\documentclass[12pt]{article}
\usepackage{gensymb,graphicx,amsmath,color,float,bm,authblk,soul,cite}
\usepackage[margin=1in]{geometry}
\graphicspath{{images/}}

\title{Actuation of cylindrical nematic elastomer balloons}
\author{Victoria Lee, Kaushik Bhattacharya}
\affil{California Institute of Technology, Pasadena CA 91125}

\begin{document} 
\maketitle

\begin{abstract}
Nematic elastomers are programmable soft materials that display large, reversible and predictable deformation under an external stimulus such as a change in temperature or light.  While much of the work in the field has focused on actuation from flat sheets, recent advances in 3D printing and other methods of directed synthesis have motivated the study of actuation of curved shells.  Snap-through buckling has been a topic of particular interest.  In this work, we present theoretical calculations to motivate another mode of actuation that combines programmable soft materials as well as instabilities associated with large deformation.  Specifically, we analyze the deformation of a cylindrical shell of a patterned nematic elastomer under pressure, show that it can undergo an enormous change of volume with changing temperature and suggest its application as a pump with extremely high ejection fraction.
\end{abstract}

\section{Introduction}

\begin{figure}
	\centering
	\includegraphics[scale=0.5]{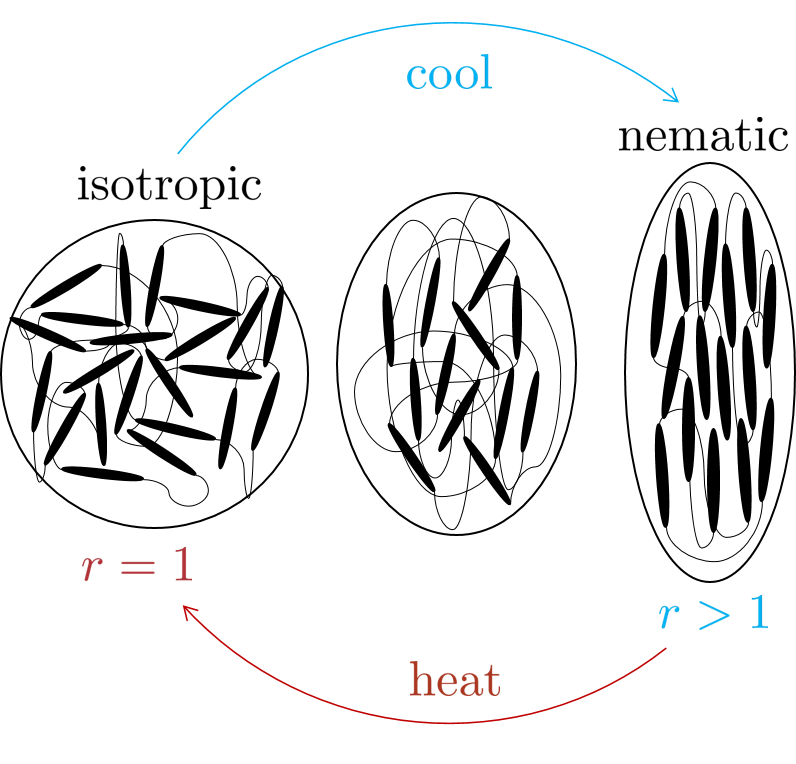}
	\caption{Isotropic-nematic phase transition in nematic elastomers with the isotropic with $r=1$ at high temperatures and nematic phase with $r>1$ at low temperatures.}
	\label{fig:LCE_schematic}
\end{figure}

Liquid crystal elastomers (LCEs) are remarkable stimuli-responsive materials that have recently been explored for their use in actuation~\cite{Warner2003}.  LCEs are elastomers with liquid crystal mesogens incorporated into the underlying polymer chains.  Of particular interest are nematic elastomers that undergo an isotropic-to-nematic transition accompanied by a significant stretch (by as much as a factor of two or more) along the nematic director and lateral contraction, as shown schematically in Figure \ref{fig:LCE_schematic}.  The stiff rod-like nematic mesogens are randomly oriented at high temperature (isotropic state) but align along a particular direction due to steric interactions below a critical temperature (nematic state).  This phase transition is induces a deformation of the underlying polymer chains in the nematic elastomer.  The degree of orientation, and consequently the change of shape, depend on temperature.

Nematic elastomers have been exploited for programmable actuation and shape-morphing of thin sheets.  Modes, Bhattacharya and Warner~\cite{Modes2010,Modes2011} suggested that if sheets of nematic elastomer with prescribed director patterns were fabricated in the nematic state and subsequently heated, they could deform out of plane into three dimensional shapes.  For example, a +1 disclination with an azimuthal director distribution would deform into a cone.  This was demonstrated in nematic glasses by de Haan \emph{et al.}~\cite{DeHaan2012}.  Ware \emph{et al.}~\cite{Ware2015} developed a method of synthesizing nematic elastomers where the director pattern could be written pixel by pixel on flat sheets and demonstrated the formation of these cones.  Moreover, they showed that this actuation was extremely robust, as the cone-lifting weights were many hundreds of times larger than the structure itself.  Since then, there have been a number of other studies on nematic elastomers~\cite{Aharoni2014,Plucinsky2016,Modes2016,Mostajeran2016,Plucinsky2018}, including the inverse problem of identifying the director pattern that would lead to a given actuated  shape~\cite{Plucinsky2016,Aharoni2018}.  All of these works address the programming and actuation of initially flat sheets.

Recent advances in 3D printing and other methods of directed synthesis have enabled the synthesis of curved shells~\cite{Gladman2016,Ambulo2017}, and such structures change shape upon heating.  In particular, Ambulo \emph{et al.} demonstrated dramatic snap-through buckling of structures involving regions of positive Gauss curvature and regions of negative Gauss curvature~\cite{Ambulo2017}.  More recently, magnetic fields have been used to independently control director orientation during 3D printing~\cite{Tabrizi2019}.    These developments in synthesis techniques motivate the current work.

The goal of this work is two-fold.  The first is to explore the combination of programmed synthesis of nematic shells and the geometric instabilities associated with the large deformation of slender structures.  Similar instabilities have been exploited in other simuli-responsive materials including electroactive materials~\cite{Goulbourne2005,Zhao2007,Rudykh2012}.  In this work, we focus on the so-called aneurysm instability of pressurized cylinders~\cite{Gent1999}.  As observed in long toy balloons, one observes a discontinuous change of radius (or volume) with an increase of pressure: typically the balloon inflates till it reaches a particular radius, beyond which point a bump (aneurysm) with a significantly larger radius appears in this region, and it propagates through the entire balloon before the radius further increases.  We explore the response of a cylindrical shell made of a nematic elastomer and study how the isotropic-nematic phase transition affects this instability.  We note here that He \emph {et al.}~\cite{He2020} have studied the anomalous behavior of (isotropic-genesis polydomain) nematic balloons under tension.  The second goal is to study actuation and shape-morphing in the presence of mechanical loads.  The prior literature has largely focused on free recovery.  

We introduce the model of the nematic elastomers at large deformation in Section \ref{sec:material_models} and analyze the deformation of a nematic elastomer cylinder under internal pressure in Section \ref{sec:cyl_twist}.   We then use the results to motivate a pump with extremely large ejection fraction in Section \ref{sec:cyl_pump}.

\section{Large deformation model of nematic elastomers}\label{sec:material_models}

We begin with the neo-classical theory of nematic elastomers following Bladon, Terentjev and Warner~\cite{Bladon1993,Warner2003}.  The state of a liquid crystal elastomer is described by an anisotropy parameter $r$, a director ${\bm n}$ and the deformation gradient ${\bm F}$ relative to a stress-free reference configuration with anisotropy parameter $r_0$ and director ${\bm n}_0$.    The anisotropy parameter is a function of temperature with $r=1$ in the isotropic state above the transformation, and gradually increases with decreasing temperature so that $r>1$ in the nematic state.  We consider the material to be incompressible so that $\det \bm{F} =1$.
The neo-classical theory considers the entropy of the polymer chains in the Gaussian approximation, and the free energy density is given as
\begin{align}
	W_{\text{WT}}(\bm{F},\bm{n},r)&=\frac{\mu}{2}\left(\text{tr}\left( \bm{\ell}_{n_0}  \bm{F}^T \bm{\ell}_{n}^{-1} \bm{F} \right)-3\right),
\end{align}
where $\mu$ is the shear modulus of the material and 
\begin{align}
\bm{\ell}_n&=r^{-1/3}\left(\bm{I}+(r-1)\bm{n}\otimes\bm{n}\right)\\
\bm{\ell}_{n_0}&=r_0^{-1/3}\left(\bm{I}+(r_0-1)\bm{n}_0\otimes\bm{n}_0\right)
\end{align}
are the step-length tensors in the current and reference configurations that collect the anisotropy parameter and the director.  It is easy to show that 
\begin{align} \label{eq:wtnh}
	W_{\text{WT}}(\bm{F},\bm{n},r)= W_{\text{NH}}\left(\bm{\ell}_{n}^{-1/2}\bm{F}\bm{\ell}_{n_0}^{1/2}\right),
\end{align} 
where $W_{\text{NH}}(\bm{F})=\frac{\mu}{2}(\text{tr }\bm{C}(\bm{F}))-3)$, with $\bm{C}(\bm{F}) = \bm{F}^T\bm{F}$, is the neo-Hookean energy density which describes the entropy of polymer chains in ordinary rubber in the Gaussian approximation~\cite{Treloar1975}.

The neo-classical theory is known to describe complex features of nematic elastomers at finite, but moderate, deformation.  However,  at extremely large stretches, the Gaussian approximation does not hold, and this theory does not adequately describe the stiffening much like its neo-Hookean counter-part.  Various constitutive relations are used to describe rubber in this high-stretch regime.  A common feature of many of these models is that the energy density depends only on principle stretches $\lambda_i$ of $\bm{F}$ (equivalently the eigenvalues $\lambda_i^2$ of  $\bm{C}(\bm{F})$):
\begin{equation} \label{eq:elastic}
W_{\text{E}}(\bm{F}) = f (\lambda_1, \lambda_2, \lambda_3).
\end{equation}
For example, in the  Ogden model~\cite{Ogden1984}  the energy density is
\begin{equation}
	W_{\text{O}}(\bm{F})=\sum_{p=1}^{N}\frac{\mu_p}{\alpha_p}\left(\lambda_1^{\alpha_p}+\lambda_2^{\alpha_p}+\lambda_3^{\alpha_p}-3\right),
\end{equation}
where $N$, $\mu_p$, and $\alpha_p$ are material constants. The shear modulus is $\mu=\frac{1}{2}\sum_{p=1}^N\mu_p\alpha_p$. When $N=1$ and $\alpha=2$, the Ogden energy is the Neo-Hookean energy, and when $N=2$, $\alpha_1=2$, and $\alpha_2=-2$, the Ogden energy is the Mooney-Rivlin energy.   We use the Ogden energy to demonstrate our results, though we can adapt them to any constitutive relation that describes the high stretch behavior.    We adopt the elastic energy density (\ref{eq:elastic}) to nematic elastomers analogously to (\ref{eq:wtnh}).   See~\cite{Agostiniani2012} for similar energies and their relaxation in the ideal case.

Further, the cross-link density and the polymer network may carry an imprint of the initial director and this leads to an additional non-ideal energy density~\cite{Biggins2012}:
\begin{equation} 
	W_{\text{NI}}(\bm{F},\bm{n})=\alpha \frac{\mu}{2}\text{tr}\left(\bm{F}(\bm{I}-\bm{n}_0\otimes\bm{n}_0)\bm{F}^T\bm{n}\otimes\bm{n}\right).
\end{equation}
Putting these together, we take the energy density of the nematic elastomer to be
\begin{equation}\label{eq:niLCEOg}
	W(\bm{F},\bm{n},r) = W_{\text{E}}\left(\bm{\ell}_{n}^{-1/2}\bm{F}\bm{\ell}_{n_0}^{1/2}\right)
	+W_{\text{NI}}(\bm{F},\bm{n}).
\end{equation}

For future use, we note a particular invariance of this energy density.  Let $\bm{Q}$ be a rotation tensor that leaves the reference director invariant: $\bm{Q}\bm{n}_0 = \pm \bm{n}_0$.  Then, we claim that 
\begin{align} \label{eq:inv}
W(\bm{Q}\bm{F}\bm{Q}^T,\bm{Q}\bm{n},r) = W(\bm{F},\bm{n},r). 
\end{align}
Note that 
\begin{equation}
\begin{aligned}
\bm{C}((\bm{Q}\bm{\ell}_{n}\bm{Q}^T)^{-1/2} (\bm{Q}\bm{F}\bm{Q}^T)\bm{\ell}_{n_0}^{1/2}) 
&= \bm{\ell}_{n_0}^{1/2}  (\bm{Q}\bm{F}\bm{Q}^T)^T (\bm{Q}\bm{\ell}_{n}\bm{Q}^T)^{-1} (\bm{Q}\bm{F}\bm{Q}^T) \bm{\ell}_{n_0}^{1/2} \\
&=\bm{\ell}_{n_0}^{1/2}  \bm{Q} \bm{F}^T \bm{\ell}_{n}^{-1} \bm{F} \bm{Q}^T \bm{\ell}_{n_0}^{1/2}\\
&= \bm{Q} \bm{\ell}_{n_0}^{1/2}  \bm{F}^T \bm{\ell}_{n}^{-1} \bm{F}  \bm{\ell}_{n_0}^{1/2} \bm{Q}^T\\
&=\bm{Q} \bm{C}( \bm{\ell}_{n}^{-1/2}\bm{F}\bm{\ell}_{n_0}^{1/2}) \bm{Q}^T,
\end{aligned}
\end{equation}
where we have used the invariance of $\bm{n}_0$ under $\bm{Q}$ in the third equality.  It follows that both tensors have the same eigenvalues and thus the same Ogden energy density.  A similar calculation holds for the non-ideal energy density as well, thereby establishing (\ref{eq:inv}).

\section{Inflation of a nematic cylinder}\label{sec:cyl_twist}



\begin{figure}
\centering
\includegraphics[width=5in]{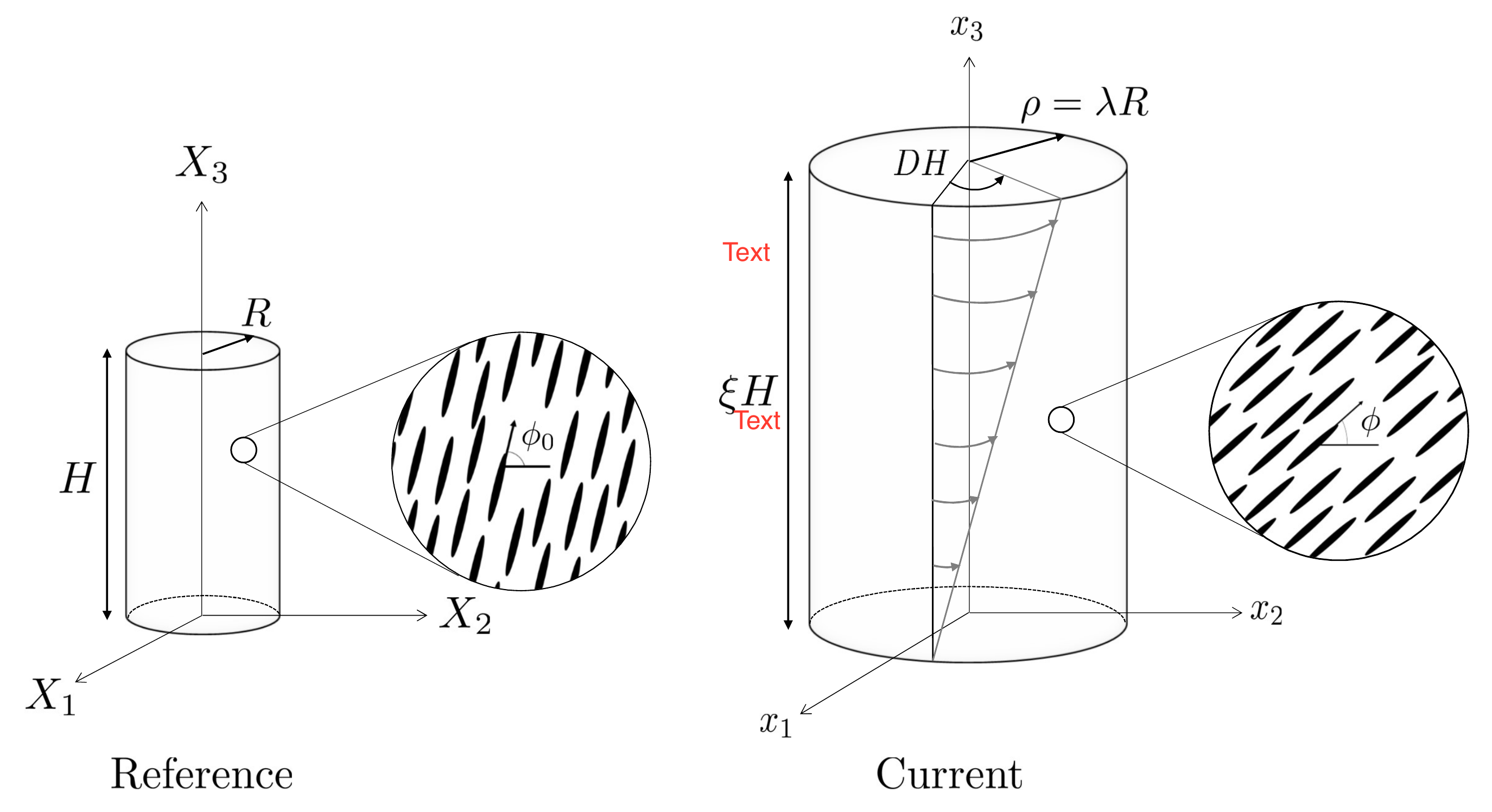}
\caption{Inflation, extension and torsion of a cylinder}
 \label{fig:def}
\end{figure}

Consider a cylindrical shell of initial (reference) length $H$, inner radius $R_i$, and undeformed outer radius $R_o$ subjected to an internal pressure $p$.  Following Rivlin~\cite{Rivlin1949} and Ericksen~\cite{Ericksen1954},  the deformation of the cylinder is described by the universal volume-preserving deformation involving radial expansion, axial extension and torsion (see Figure \ref{fig:def}). The mapping is
%
%
%
\begin{equation}
	\begin{cases}
		\rho=\rho(R)\\
		\theta=\Theta+DZ\\
		z=\xi Z
	\end{cases},
\end{equation}
where $\{R,\Theta,Z\}$ and $\{\rho,\theta,z\}$ denote the cylindrical coordinate system in the reference and deformed coordinate systems respectively. $\rho(R)$ describes the radial expansion, $D$ the twist and $\xi$ the axial stretch.   The deformation gradient in the cylindrical coordinate system is 
\begin{equation} \label{eq:f}
\bm{F}
=	\begin{pmatrix}
		\rho' && 0 && 0 \\
		0 && {\frac{\rho}{ R}} && \lambda R D \\
		0 && 0 && \xi
	\end{pmatrix} 
=	\begin{pmatrix}
		\frac{1}{\lambda\xi} && 0 && 0 \\
		0 && \lambda && \lambda R D \\
		0 && 0 && \xi
	\end{pmatrix},
\end{equation}
where we have introduced the hoop stretch $\lambda(R) = \rho(R)/R$ and used the incompressibility to obtain the second equality.

We assume that the director both in the reference and deformed configuration are tangential to the cylinder and make an angle $\phi_0$ and $\phi$, respectively, with the azimuthal coordinate.  Thus, in cylindrical coordinates,
\begin{equation} \label{eq:n0n}
\bm{n}_0=
	\begin{pmatrix}
		0 \\
		\cos \phi_0 \\
		\sin \phi_0
	\end{pmatrix}
\text{ and } \bm{n}=
	\begin{pmatrix}
		0 \\
		\cos \phi \\
		\sin \phi
	\end{pmatrix}.
\end{equation}
The total potential energy of the system is
\begin{equation}
\Phi=\int_{\Omega}{W(\bm{F},\bm{n},r)  dV}-p\Delta V,
\end{equation}
where $\Delta V$ is the difference in the deformed and undeformed volumes. Applied to a balloon with height $H$, we obtain
\begin{equation}
\begin{aligned}
\Phi&=\int_{0}^{H}{\int_{0}^{2\pi}{\int_{R_i}^{R_o}{W(\bm{F},\bm{n},r) R dR}d\Theta} dZ}-p\left(\pi \rho^2 \xi H-\pi R^2 H\right)\bigg\rvert_{R=R_i}\\
&\approx2\pi R_i H T W(\bm{F},\bm{n},r) - p\pi R_i^2 H\left( \xi \lambda^2-1\right). \label{eq:cylexpenergy}
\end{aligned}
\end{equation}
Above we have assumed that the shell is thin, $T:=(R_o-R_i)<<R_i$, to evaluate the integral.

For a given pressure $p$ and anisotropy parameter $r$, we can now find the equilibrium as 
\begin{equation} \label{eq:eq}
	\frac{\partial{\Phi}}{\partial{\lambda}}=\frac{\partial{\Phi}}{\partial{\xi}}=\frac{\partial{\Phi}}{\partial{D}}=\frac{\partial{\Phi}}{\partial{\phi}}=0.
\end{equation}
Physically, these equations describe the balance between the hoop stress in the cylinder and the internal pressure, the balance between the axial stress and the internal pressure, the balance of torque, and the balance of internal (material) torque on the director respectively. 

\begin{table} \label{tab:par}
\centering
\caption{Table of parameters}
\begin{tabular}{lll}
    Inner radius & $R_i$ & 1 cm\\
    Outer radius & $R_o$ & 1.05cm\\
    Height of cylinder & $H$ & 5cm \\
    Initial director angle & $\phi_0$ & $90\degree$\\
    Initial anisotropy parameter & $r_0$ & 2\\
    Non-ideality parameter & $\alpha$ & 0.3 \\
    Ogden model shear modulus & $\mu_1$ &  $1.0 \cdot 10^5$ Pa \\
    Ogden model shear modulus & $\mu_2$ &  $1.904762\cdot 10^2$ Pa \\
    Ogden model shear modulus & $\mu_3$ &  $-1.5873\cdot 10^3$ Pa \\
    Ogden model constant & $\alpha_1$ & 1.3\\
    Ogden model constant & $\alpha_2$ & 6\\
    Ogden model constant & $\alpha_3$ & -3
 \end{tabular}
\end{table}

\begin{figure}
\centering
\includegraphics[width=5in]{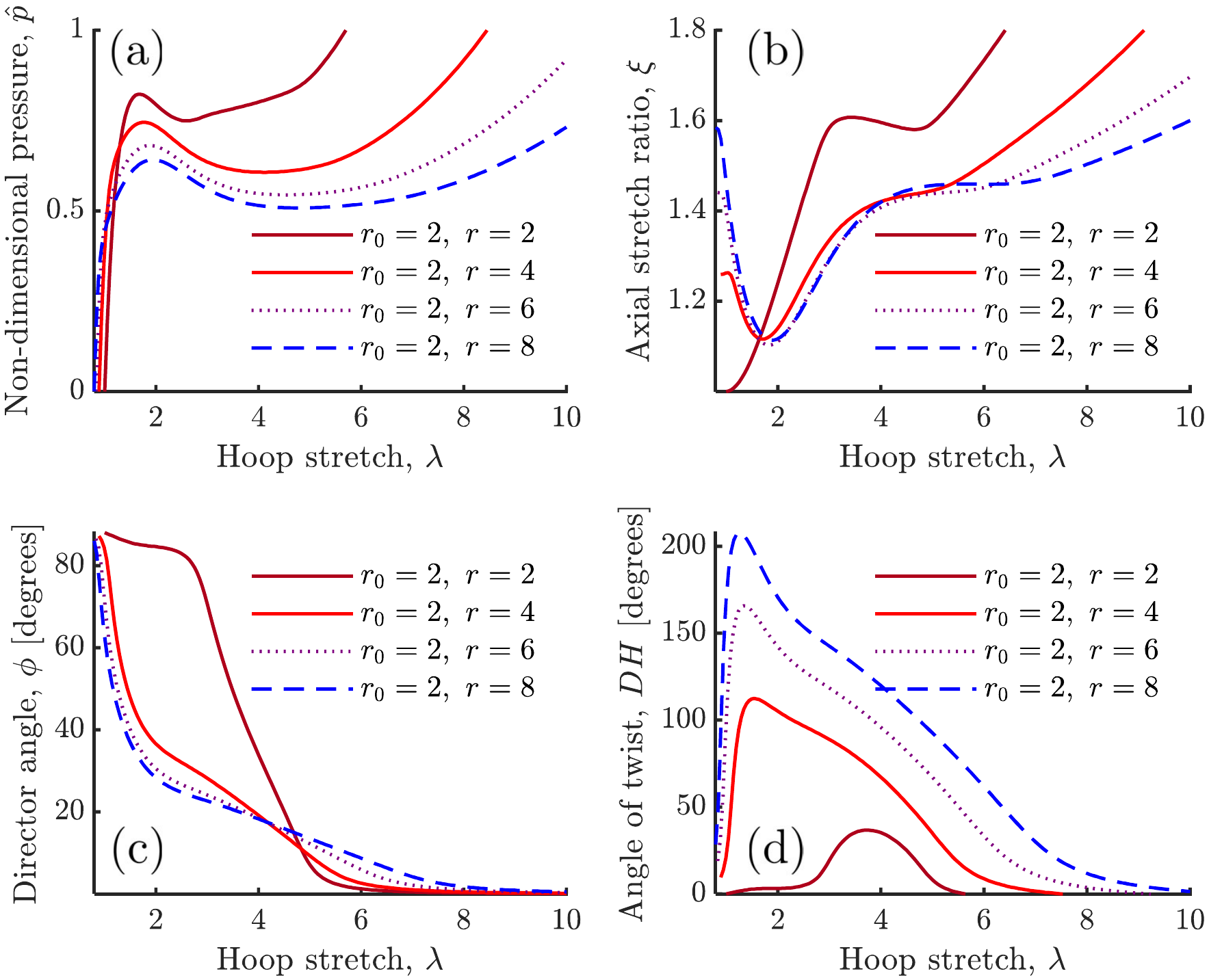}
\caption{Inflation of a nematic cylinder: Solution 1. (a) Pressure vs. hoop stretch, (b) Axial stretch  vs. hoop stretch, (c) Director angle  vs. hoop stretch,
(d) Twist  vs. hoop stretch}
\label{fig:sol1}
\end{figure}

\begin{figure}[t]
\centering
\includegraphics[width=5in]{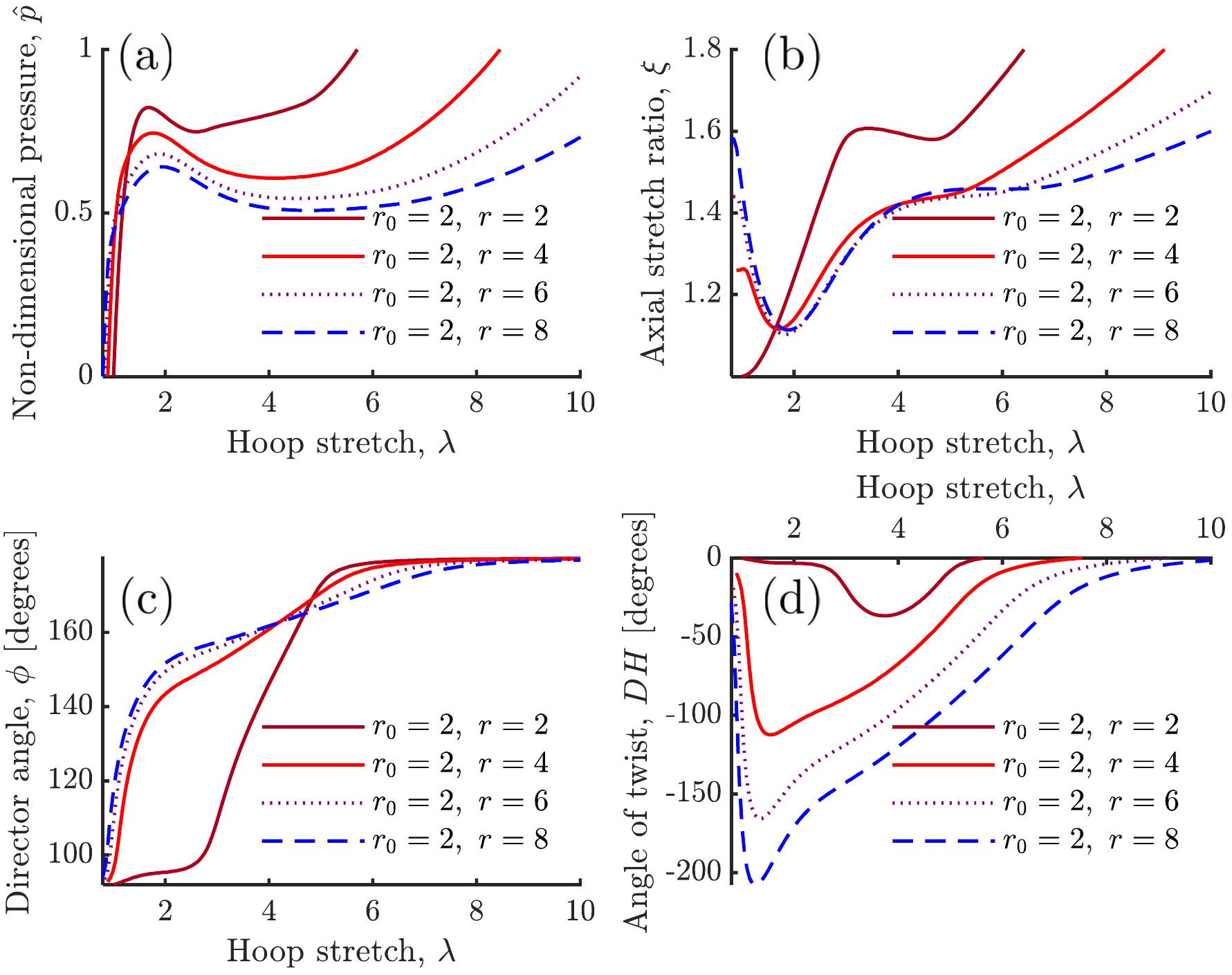}
\caption{Inflation of a nematic cylinder: Solution 2. (a) Pressure vs. hoop stretch, (b) Axial stretch  vs. hoop stretch, (c) Director angle  vs. hoop stretch,
(d) Twist  vs. hoop stretch}
 \label{fig:sol2}
\end{figure}

To demonstrate the results, we consider a cylinder where the initial director is axial ($\phi_0 = 90\degree$) and which is mildly nematic with initial anisotropy parameter $r_0 = 2$.  The rest of the parameters are shown in Table \ref{tab:par}.  We fix the current anisotropy parameter $r$ and the hoop stretch $\lambda$ and solve (\ref{eq:eq}) for the pressure $p$, axial stretch $\xi$, the twist $D$ and the current director angle $\phi$.  We find that the system has two solutions, shown in Figures \ref{fig:sol1} and \ref{fig:sol2} for four different current anisotropy parameters $r$\footnote{There is  a third unstable solution where the director does not rotate which we ignore.}.

Consider the first solution, Figure \ref{fig:sol1}.  The hoop stretch vs. pressure is non-monotone (Figure \ref{fig:sol1}(a)): the pressure initially increases but then drops before increasing again with increasing hoop stretch.  This reflects the well-known balloon instability: with increasing pressure, the radius would increase till it reaches a critical pressure at which it will jump to a large radius.  Note further that the critical pressure decreases and the change of hoop stretch increases with increasing anisotropy parameter $r$.  The axial stretch decreases initially but increases with the onset of the instability (Figure \ref{fig:sol1}(b)); this is consistent with the behavior of rubber balloons.  Figure \ref{fig:sol1}(c) shows that the director begins to rotate with inflation, reaching the hoop direction asymptotically.  To understand this, an increase in radius increases the volume more than an increase in axial stretch.  Therefore, the pressure seeks to increase the circumference by reorienting the director.  This is resisted by the non-ideality, and the balance leads to the observed behavior; so the reorientation would increase with decreasing non-ideality constant $\alpha$ and vice-versa.  This reorientation leads to a twist in the cylinder.

\begin{figure}[t] 
\centering
\includegraphics[width=4in]{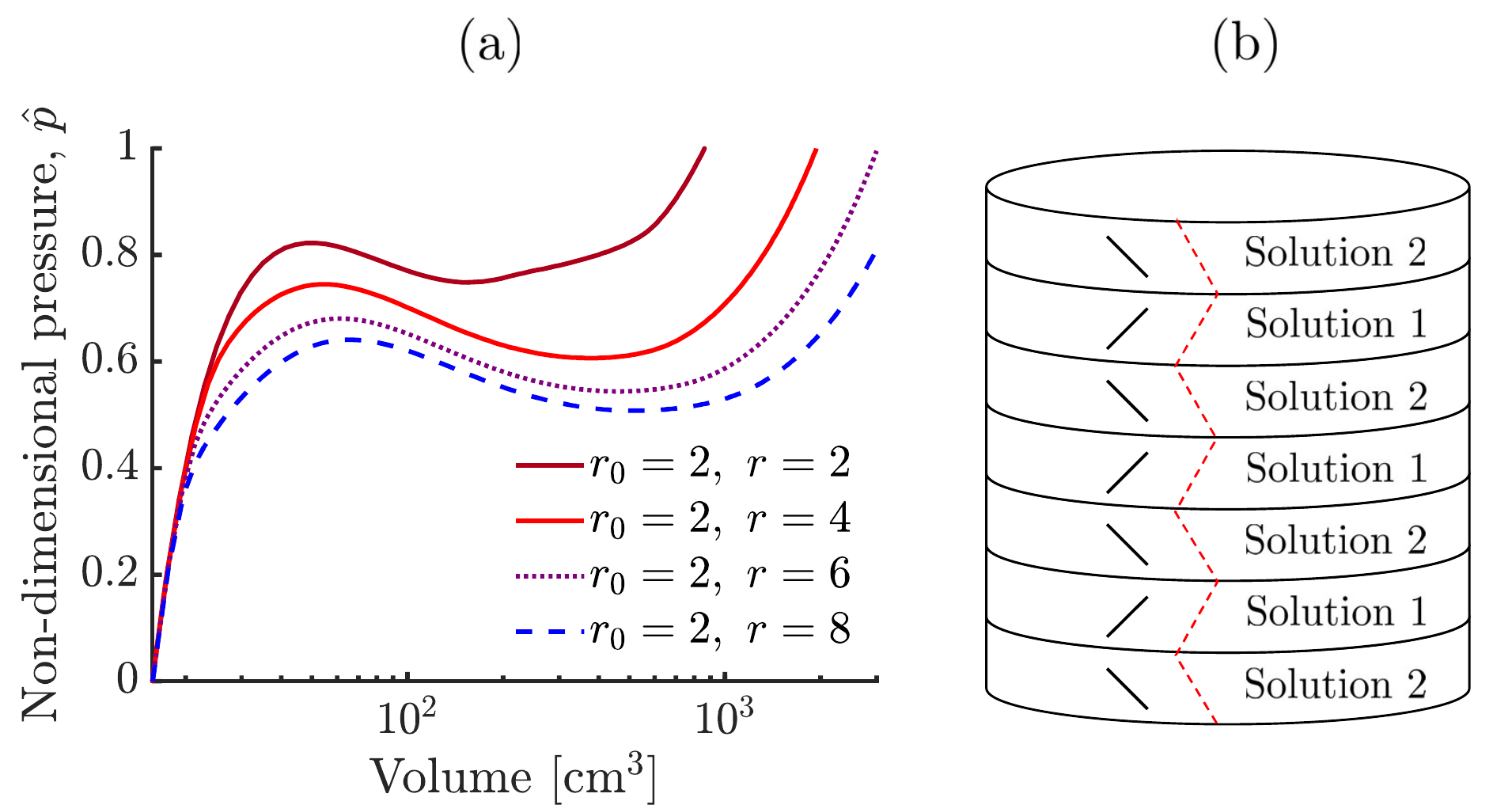}
\caption{(a) Pressure-volume relation for the inflation of a nematic cylinder. (b) Formation of stripe domains that avoid any torsion -- the lines describe the director, while the dashed line is the deformed shape of a fiducial line that is initially straight and axial.}
\label{fig:pvstripe}
\end{figure}

The second solution, Figure \ref{fig:sol2}, is very similar to the first, except that the reorientation and twist change sign.  The pressure vs. hoop-stretch and the axial stretch vs. hoop-stretch curves remain unchanged.  Consequently, both solutions have the same pressure vs. volume curves which are shown in Figure \ref{fig:pvstripe}(a).  The volume is plotted on a logarithmic scale due to the dramatic change of volume during the instability.  

It is useful to understand the origin of the two solutions.  The material is not chiral, and neither is the initial configuration.  Therefore a breaking of the chiral symmetry by rotation of the director has to be accompanied by a symmetry-related counterpart.  To elaborate on this, recall the invariance (\ref{eq:inv}).  Let ${\bf Q}$ be a  $180\degree$ rotation about the azimuthal direction,
\begin{align}
\bm{Q} = \begin{pmatrix} -1 & 0 & 0 \\ 0 & 1 & 0 \\ 0 & 0 &-1 \end{pmatrix}.
\end{align}
Note that $\bm{Q} \bm{n}_0 = -\bm{n}_0$ so that it satisfies the requirement for (\ref{eq:inv}).  It is easy to check that for $\bm{F}$ and $\bm{n}$ in (\ref{eq:f}) and (\ref{eq:n0n}),
\begin{align}
\bm{Q} \bm{F} \bm{Q}^T = \begin{pmatrix}
		\frac{1}{\lambda\xi} && 0 && 0 \\
		0 && \lambda && -\lambda R D \\
		0 && 0 && \xi
	\end{pmatrix}, \quad
\bm{Q} \bm{n}=
	\begin{pmatrix}
		0 \\
		\cos \phi \\
		- \sin \phi
	\end{pmatrix}.	
\end{align}
Thus, the invariance  (\ref{eq:inv}) implies that any solution with chirality has a symmetric counterpart with the same radial and azimuthal stretches.

The presence of the two symmetric solutions enables the formation of stripe domains that avoid overall torsion as shown in Figure \ref{fig:pvstripe}(b).  
We divide the cylinder into short cylindrical rings and alternate between the two solutions.  This leads to a continuous deformation, where one ring twists one way and the other the other way in an alternating pattern, but they meet continuously across the boundaries as indicated by the initially straight fiducial dashed line shown in the figure.  The overall torsion is zero while the overall hoop and axial stretch are as before, leading to the pressure-volume curve shown in \ref{fig:pvstripe}(a).  

Stripe domains are widely observed in nematic elastomers, especially in uniaxial tension, where rigid grips prevent any shear~\cite{Warner2003}.  In uniaxial tension of a nematic sheet along an axis that is perpendicular to the initial director orientation, director rotation accommodates stretch but causes shear.  However, shear breaks the symmetry and therefore there are two solutions (rotation to the right or left), which alternate to form the stripe domains.  The domains are fine, typically with the width of microns, and the interfaces are very sharp, with a width of nanometers.  The stripe domains in Figure \ref{fig:pvstripe}(b) are the exact analogs of those in uniaxial tension.

\section{Pump }\label{sec:cyl_pump}

\begin{figure}[t]
	\centering
	\includegraphics[width=0.75\textwidth]{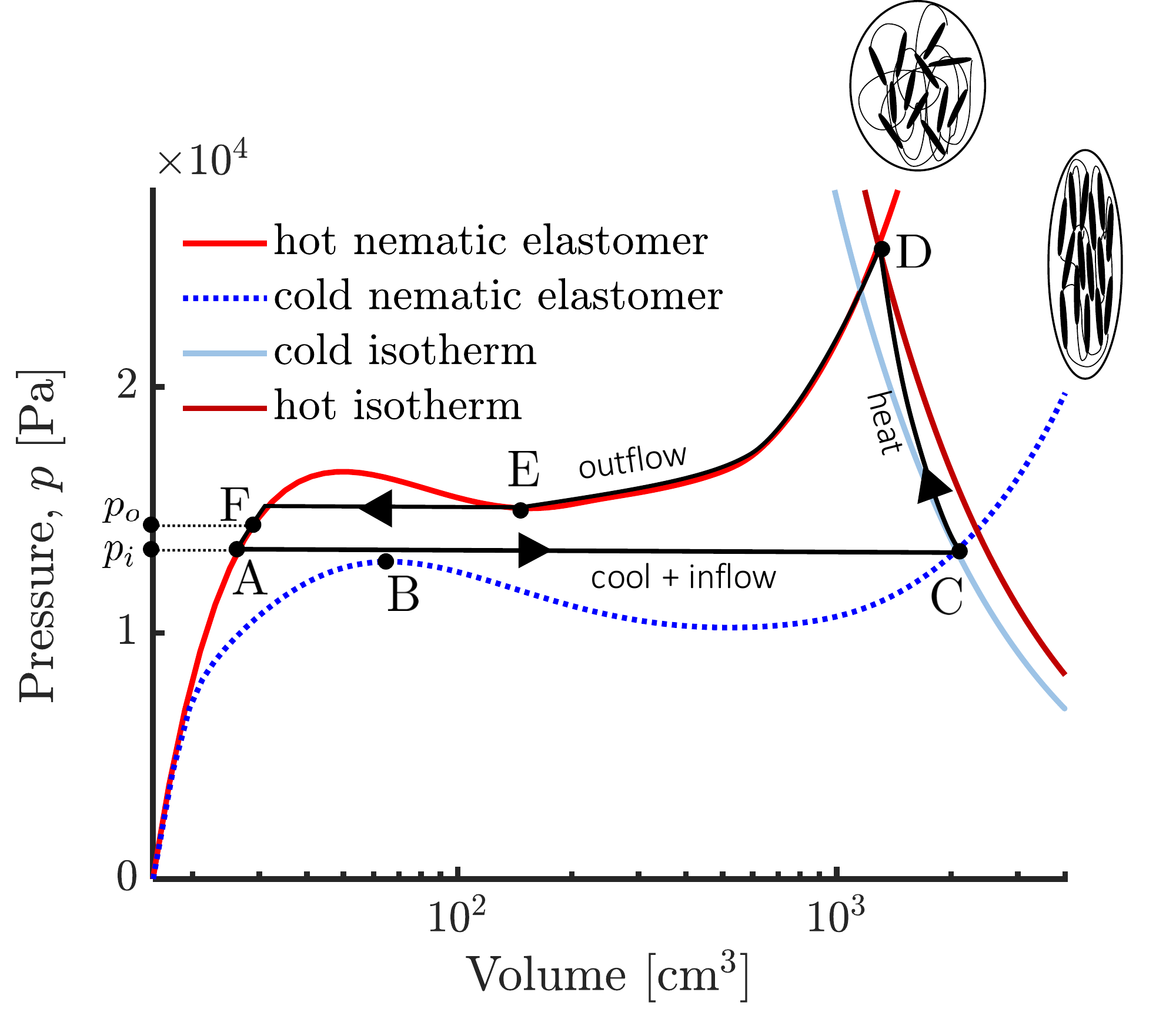}
	\caption{The operation of a pump with an input pressure $p_i$ and output pressure $p_o$ by heating and cooling a nematic cylinder.}
	\label{fig:pump}
\end{figure}

The pressure-volume curves in Figure \ref{fig:pvstripe}(a) motivate the application of this cylindrical nematic elastomer balloon as a pump.  Recall that the anisotropy parameter $r$ depends on temperature, and therefore the four pressure-volume curves represent four distinct temperatures. In a typical monodomain nematic elastomer, $r=2$ at a high temperature of about $85\degree$C, while  $r=8$ at a low temperature of about $25\degree$C~\cite{Warner2003}. These two pressure-volume curves are re-plotted in Figure \ref{fig:pump} as the hot and cold nematic elastomers.  An important observation is that the lower-critical pressure (point E) of the pressure-volume curve at the high temperature is higher than the upper-critical pressure (point B) of the pressure-volume curve at the low temperature.  This enables the operation as a pump between an inlet pressure $p_i$ and outlet pressure $p_o$, where $p_B < p_i \le p_o \le p_E$.
Also shown in the figure are the isotherms (pressure-volume relation) of a fixed mass of fluid, in this case air, at the hot and the cold temperatures of $85\degree$C and $25\degree$C respectively.

The pump operates as follows.  Let us begin at the high temperature with the outlet closed and the inlet open so that the nematic pump is at the point A.     Now, cool the pump with the inlet open.  At the inlet pressure $p_i$, which is larger than the upper critical pressure of the cold balloon, the only equilibrium solution is the point C, which has a very large volume.  Therefore, as the pump is cooled, its volume expands from A to B, drawing in fluid from the inlet.  Now close the inlet and start heating the pump.  The mass of fluid in the pump is fixed, and so its behavior shifts from that of the cold isotherm to that of the hot isotherm.  In the interim, the pressure-volume curve of the pump also changes to that of the hot material.   Therefore, the equilibrium shifts from C to D.  Now, open the outlet so that the pressure decreases to $p_o$.  The only available state in the hot pump is at F, and so the pump goes from point D, with very large volume, to point F, with small volume, expelling the fluid. Closing the outlet and opening the inlet takes us from F to A, resetting the pump.

A pump can be characterized by its ejection fraction. In this case, the ejection fraction is
\begin{equation}\label{key}
	\frac{(\text{filled volume})-(\text{empty volume})}{\text{filled volume}}=\frac{V_C-V_A}{V_C}=98.7\%,
\end{equation}
which means that $98.7\%$ of the fluid is pumped out of a filled balloon during each cycle.  This is extremely high: a normal human heart has a left ventricular ejection fraction between $50\%$ and $70\%$, according to the American College of Cardiology~\cite{Kosaraju2020}.

\section{Conclusion and future work}

We have introduced a modified formulation of the standard Warner-Terentjev energy density incorporated into a higher-order Ogden model to more accurately describe the behavior of nematic elastomers at large deformation. Furthermore, this work has initiated the study of actuation from geometries beyond flat, two-dimensional sheets by exploring a curvilinear three-dimensional geometry. We have outlined the deformation of a nematic elastomer balloon under simple expansion and twist. The material is actuated remotely by changing the temperature to dictate the degree of anisotropy, and the response is tunable. The foundation for our understanding of nematic elastomer actuation from flat geometries has already been well established with respect to the design, optimization, manufacturing, and tuning (e.g. voxelated sheets~\cite{Ware2015}, wrinkling-resistant membranes~\cite{Plucinsky2017}, and moving inchworm~\cite{Yamada2009}). Future applications based on more complex geometries and loading conditions, for instance incorporation of disclination defects and gradients of director or temperature across the thickness, can build upon this framework.

\section*{Acknowledgments}
We are grateful for the financial support of the US Air Force Office of Scientific Research through the MURI Grant No. FA9550-18-1-0566. 

\section*{Data Availability}
The data that supports the findings of this study are available within the article.

\bibliographystyle{abbrv}
\bibliography{library2}

\end{document}